\documentclass[aps,prb,preprint,groupedaddress,showpacs]{revtex4}
\usepackage{color} %for the comments
\usepackage{graphicx}
\usepackage[T1]{fontenc}
\begin{document}
%\preprint{}

\title{Experimental application of sum rules for electron energy loss magnetic chiral dichroism}

\author{L. Calmels}
\author{F. Houdellier}
\author{B. Warot-Fonrose}
\author{C. Gatel}
\author{M. J. H\"ytch}
\author{V. Serin}
\author{E. Snoeck}
\affiliation{CEMES-CNRS, 29 rue Jeanne Marvig, BP 94347, Toulouse Cedex 4, France}

\author{P. Schattschneider}
\affiliation{Institute for Solid State Physics, Vienna University of Technology,\\Wiedner Hauptstrasse 8-10/138, A-1040 Vienna, Austria.}

\date{\today}

\begin{abstract}
We present a derivation of the orbital and spin sum rules for magnetic circular dichroic spectra measured by electron energy loss spectroscopy in a transmission electron microscope. These sum rules are obtained from the differential cross section calculated for symmetric positions in the diffraction pattern. Orbital and spin magnetic moments are expressed explicitly in terms of experimental spectra and dynamical diffraction coefficients. We estimate the ratio of spin to orbital magnetic moments and discuss first experimental results for the Fe $L_{2,3}$ edge.
\end{abstract}


\maketitle

%\section{}
%\subsection{}
%\subsubsection{}
Electron energy loss spectroscopy (EELS) in a transmission electron microscope (TEM) gives access at high energy losses to the density of unoccupied valence states with a sub-nanometer spatial resolution \cite{batson93,muller99}. The possibility of using EELS to measure an energy loss magnetic chiral dichroism (EMCD) spectrum analogous to the X-ray magnetic circular dichoism (XMCD) signal obtained with synchrotron radiation \cite{schutz87,chen95} has been suggested in 2003  \cite{hebert03} and demonstrated recently \cite{schatt06}. The principles of an EMCD experiment are the following: after suitable orientation of the sample, the incident electron beam is first elastically diffracted by the crystal. Each diffracted beam is then inelastically scattered. The total inelastic signal can be written as the sum of two kinds of contributions: the first one is due to each single diffracted beam and can be written in terms of the dynamic form factors (DFF) $S\left(\textbf{q},\textbf{q},E\right)$, where $\textbf{q}$ is the momentum transfer which depends on the diffracted beam and on the location of the spectrometer aperture in the diffraction pattern, and $E$ is the energy loss. The second contribution involves all the possible pairs of diffracted beams and is described by the mixed dynamic form factors (MDFF) $S\left(\textbf{q},\textbf{q}',E\right)$ \cite{schatt96}.

\begin{figure}[h]
\begin{center}
\includegraphics[scale=0.6]{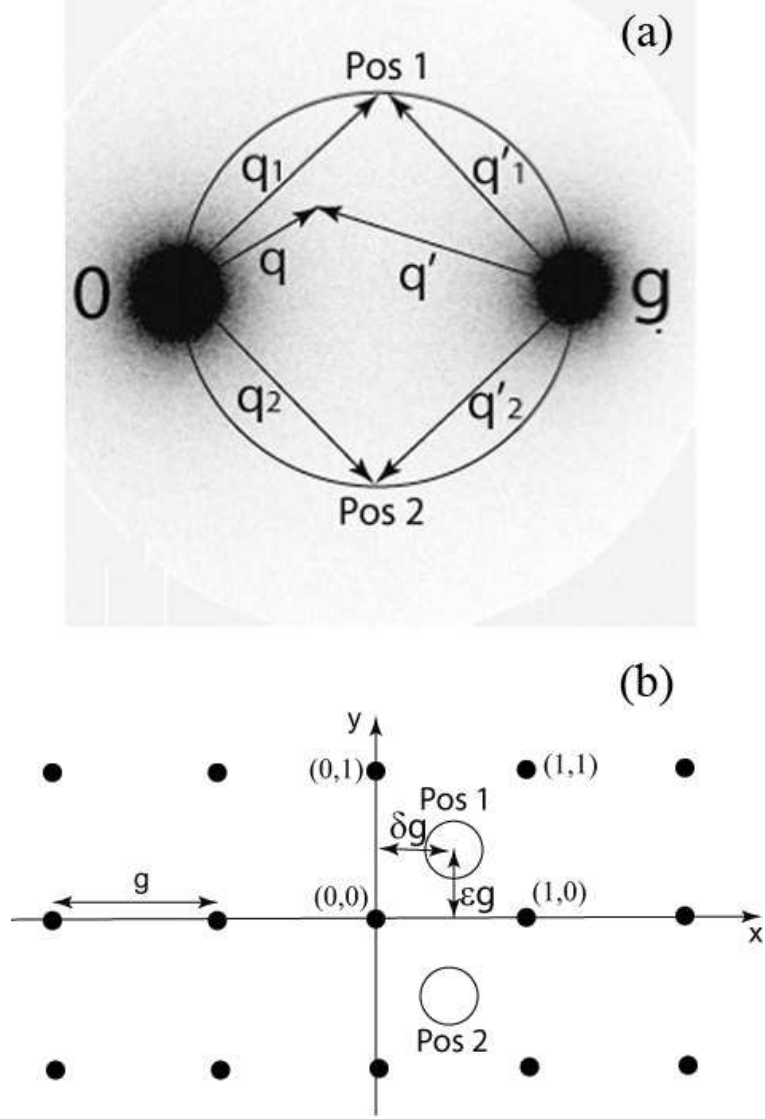}
\caption{\label{fig:diffraction}Diffraction pattern for an EMCD experiment. (a): experimental diffraction pattern for an iron sample oriented in (110) two beam configuration. The transfer momentum vectors for the two symmetrical positions pos1 and pos2 are represented by arrows. (b): Four fold diffraction pattern which has been used to express the differential cross section. The two different positions of the spectrometer aperture ($\textbf{OS}=\delta g\textbf{e}_{x}\pm\epsilon g\textbf{e}_{y}$, with $\delta$ and $\epsilon$ real numbers) which have been considered are indicated by open circles. The Bragg spots are represented by filled circles. A pair of integers $\left(n,m\right)$ is associated to each Bragg spot, as shown for four of them.}
\end{center}
\end{figure}

Inelastic scattering events are due to the coulomb interaction between the electrons of the probe and the sample electrons. In quantum electrodynamics, this interaction is described in terms of a virtual photon exchanged between the two electrons. The virtual photon associated to one of the DFF is linearly polarized in the direction of the transfer momentum vector (Lorentz Gauge). The polarization of the virtual photon associated to one of the MDFF is more complicated for any couple of transfer momentum vectors $\textbf{q}$ and $\textbf{q}'$. It becomes right or left circularly polarized when the phase difference between the diffracted beams is $\pi/2$, and when the momentum transfer vectors $\textbf{q}$ and $\textbf{q}'$ are orthogonal with identical modulus. This is the case for the vectors $\left\{\textbf{q}_{1},\textbf{q}'_{1}\right\}$ and $\left\{\textbf{q}_{2},\textbf{q}'_{2}\right\}$ shown in Fig.~\ref{fig:diffraction}a for which $\left(\textbf{q}_{1},\textbf{q}'_{1}\right)=\pi/2$ and $\left(\textbf{q}_{2},\textbf{q}'_{2}\right)=-\pi/2$.
The EMCD signal is in this case obtained by subtracting the spectra measured at the two positions pos1 and pos2 shown in Fig.~\ref{fig:diffraction}a. An accurate description of the EMCD spectra is not trivial, firstly because all the pairs of diffracted beams must be considered together, secondly because the propagation of the diffracted beams must be described within the fast electron dynamical diffraction theory, the incident and scattered electron beams behaving like Bloch waves inside the crystal \cite{spence92}.

Magnetic circular dichroism has been measured in a TEM on the $L_{2,3}$ edges of 3$d$ magnetic metals \cite{schatt06}. The most recent experimental papers describe the configurations which give the highest dichroic signal as well as dichroic/noise ratio. Several configurations have been tested to reach this aim, like using convergent instead of parallel incident beam to increase the total current (LACDIF configuration \cite{midgley99,morniroli07,warot07}), or choosing the sample orientation and searching for the positions in the diffraction pattern which enhance the dichroic signal. This experimental investigation has been done by moving the diffraction pattern over the spectrometer aperture \cite{schatt06}, or with the energy spectrum imaging technique (ESI), which consists in recording the whole diffraction pattern for successive energy windows of typically 1 eV running over the $L_{2,3}$ edges \cite{warot07}. The EMCD signal has also been calculated theoretically. These calculations, which are based on the first principles determination of the fast electron Bloch wave functions \cite{spence92,schatt96} and transition matrix elements \cite{schatt06,rusz07} have been very helpful to find the experimental conditions and sample characteristics which give the highest dichroic signal.

Up to now, EMCD experiments have only been analysed quantitatively in terms of dichroism. A quantitative interpretation of the spectra requires the determination of new sum rules which take into account the dynamical diffraction effects. In this paper, we present the analytical derivation of the orbital and spin sum rules for magnetic chiral dichroic spectra measured by EELS, and we discuss to which extent these sum rules can be applied to experimental results.

The EELS spectra measured in a TEM can be described by the differential cross section $\frac{\partial^{2}\sigma}{\partial E\partial\Omega}$ for scattering of a fast probe electron with the energy loss $E$ and scattering angle $\Omega$. When the spectrometer aperture is located at a given position in the $\left\{x,y\right\}$ plane of the diffraction pattern, the differential cross section for the core electron excitation edges can be written as \cite{hebert03}
\begin{equation}
\frac{\partial^{2}\sigma}{\partial E\partial\Omega}=\sum_{\textbf{q}}\frac{A^
{det}_{\textbf{q};\textbf{q}}}{q^{4}}S\left(\textbf{q},\textbf{q},E\right)+\sum_{\textbf{q}}\sum_{\textbf{q}'\neq\textbf{q}}2Re\left[\frac{A^{det}_{\textbf{q};\textbf{q}'}}{q^{2}q'^{2}}S\left(\textbf{q},\textbf{q}',E\right)\right]\label{eq:one},
\end{equation}
where the mixed dynamic form factors of one atom are given by
\begin{equation}
S\left(\textbf{q},\textbf{q}',E\right)=\sum_{i,f}\left\langle i\right|\textrm{exp}\left(-i\textbf{q}\cdot\textbf{r}\right)\left|f\right\rangle\left\langle f\right|\textrm{exp}\left(i\textbf{q}'\cdot\textbf{r}\right)\left|i\right\rangle\delta\left(E-E_{f}+E_{i}\right).
\end{equation}
$\left|i\right\rangle$ and $\left|f\right\rangle$ are the initial core states and the final unoccupied valence states with energies $E_i$ and $E_f$. $\textbf{q}=\textbf{OS}-\textbf{g}+q_{z}\textbf{e}_{z}$ and $\textbf{q}'=\textbf{OS}-\textbf{g}'+q_{z}\textbf{e}_{z}$ are the momentum transfer vectors which depend on the vector $\textbf{OS}$ connecting the transmitted beam and the spectrometer aperture in the diffraction pattern, on the reciprocal lattice vectors $\textbf{g}$ and $\textbf{g}'$, and on the momentum $q_{z}<0$ which is transferred in the incident beam direction $\left(Oz\right)$. The first and second terms in the right hand side of Eq.~(\ref{eq:one}) describe respectively the contributions from the DFF and MDFF. The double sum over $\textbf{q}$ and $\textbf{q}'$ implies that the pairs of Bragg spots are not counted twice. The coefficients $A^{det}_{\textbf{q};\textbf{q}'}$ contain all the information on the Bloch wave eigenvectors and eigenvalues which is needed to describe the fast incident and scattered electron beams within the framework of the dynamical diffraction theory. They can be calculated as shown recently \cite{rusz07}. They depend on the momentum transfers $\textbf{q}$ and $\textbf{q}'$, on the location of the spectrometer aperture, on the atomic structure of the crystal, on the thickness and orientation of the sample, and on the location of the ionized atom inside the sample. For more complex unit cells, Eq.~(\ref{eq:one}) needs to be generalized by summing over the different atom species. The operators $r_{+}=x+iy$, $r_{-}=x-iy$, and $r_{0}=z$ can be used to express the mixed dynamic form factors within the electric dipole approximation as
\begin{equation}
S\left(\textbf{q},\textbf{q}',E\right)=\frac{q_{x}q'_{x}+q_{y}q'_{y}}{4}\left
(\mu_{+}+\mu_{-}\right)+q^{2}_{z}\mu_{0}+i\frac{q_{x}q'_{y}-q_{y}q'_
{x}}{4}\left(\mu_{+}-\mu_{-}\right)
\end{equation}
where
\begin{equation}
\mu_{+}=\sum_{i,f}\left|\left\langle i\right|r_{+}\left|f\right\rangle\right|^
{2}\delta\left(E-E_{f}+E_{i}\right),
\end{equation}

\begin{equation}
\mu_{-}=\sum_{i,f}\left|\left\langle i\right|r_{-}\left|f\right\rangle\right|^
{2}\delta\left(E-E_{f}+E_{i}\right),
\end{equation}
and
\begin{equation}
\mu_{0}=\sum_{i,f}\left|\left\langle i\right|r_{0}\left|f\right\rangle\right|^
{2}\delta\left(E-E_{f}+E_{i}\right),
\end{equation}
$z$ being the quantization axis. In the following, we have considered a four-fold diffraction pattern with distance $g$ between Bragg spots and the two spectrometer aperture positions pos1 and pos2 which are indicated in Fig.~\ref{fig:diffraction}b. Such a diffraction pattern can be observed with \textit{bcc} Fe or \textit{fcc} Ni crystals oriented in the (100) zone axis. The reciprocal lattice vectors are given by $\textbf{g}=ng\textbf{e}_{x}+mg\textbf{e}_{y}$ ($n$ and $m$ being the integers associated to each Bragg spot), and the momentum transfer vectors are written as $\textbf{q}=g\left(\delta-n\right)\textbf{e}_{x}+g\left(\epsilon-m\right)\textbf{e}_{y}+q_{z}\textbf{e}_{z}$ for position 1, and by $\textbf{q}=g\left(\delta-n\right)\textbf{e}_{x}-g\left(\epsilon+m\right)\textbf{e}_{y}+q_{z}\textbf{e}_{z}$ for position 2 ($\delta$ and  $\epsilon>0$ being real). The difference and the sum between the EELS signals measured at the two symmetric positions of the spectrometer aperture described above are given by
$$
\sigma_{2}\mp\sigma_{1}=
$$
$$
 \frac{1}{4g^{2}} 
\sum_{\left(n,m\right)}
\frac{A^{pos2}_{n,-m;n,-m}\mp A^{pos1}_{n,m;n,m}}
{\left[\left(\delta-n\right)^{2}+\left(\epsilon-m\right)^{2}+
\frac{q^{2}_{z}}{g^{2}}\right]^{2}}
\left\{\left[\left(\delta-n\right)^{2}+\left(\epsilon-m\right)^{2}\right]\left
(\mu_{+}+\mu_{-}\right)+4\frac{q^{2}_{z}}{g^{2}}\mu_{0}\right\}
$$
$$
+\frac{1}{2g^{2}}
\sum_{\left(n,m\right)}\sum_{\left(n',m'\right)\neq\left(n,m\right)}
\frac{\textrm{Re}\left(A^{pos2}_{n,-m;n',-m'}\mp A^{pos1}_{n,m;n',m'}\right)}
{\left[\left(\delta-n\right)^{2}+\left(\epsilon-m\right)^{2}+\frac{q^{2}_{z}}{g^
{2}}\right]
\left[\left(\delta-n'\right)^{2}+\left(\epsilon-m'\right)^{2}+\frac{q^{2}_{z}}
{g^{2}}\right]}
$$
$$
\times\left\{\left[\left(\delta-n\right)\left(\delta-n'\right)+\left(\epsilon-
m\right)\left(\epsilon-m'\right)\right]\left(\mu_{+}+\mu_{-}\right)+4\frac{q^{2}
_{z}}{g^{2}}\mu_{0}\right\}
$$
$$
+\frac{1}{2g^{2}}
\sum_{\left(n,m\right)}\sum_{\left(n',m'\right)\neq\left(n,m\right)}
\frac{\textrm{Im}\left(A^{pos2}_{n,-m;n',-m'}\pm A^{pos1}_{n,m;n',m'}\right)}{\left[\left
(\delta-n\right)^{2}+\left(\epsilon-m\right)^{2}+\frac{q^{2}_{z}}{g^{2}}\right]
\left[\left(\delta-n'\right)^{2}+\left(\epsilon-m'\right)^{2}+\frac{q^{2}_{z}}
{g^{2}}\right]}
$$
\begin{equation}
\times\left[\left
(\delta-n\right)\left(\epsilon-m'\right)-\left(\delta-n'\right)\left(\epsilon-m\right)\right]
\left(\mu_{+}-\mu_{-}\right)\label{eq:seven},
\end{equation}
where $\sigma_{2}=\left(\frac{\partial^{2}\sigma}{\partial E\partial\Omega}\right)_{pos2}$, $\sigma_{1}=\left(\frac{\partial^{2}\sigma}{\partial E\partial\Omega}\right)_{pos1}$, and the pairs of Bragg spots $\left(n,m\right)$ and $\left(n',m'\right)$ are not counted twice in the double sum. In the perfect zone axis configuration, this equation can be simplified using
\begin{equation}
A^{pos1}_{n,m;n',m'}=A^{pos2}_{n,-m;n',-m'}\label{eq:eight}.
\end{equation}
These equations remain valid in the systematic row configuration which is reached by tilting the sample around the $\left(Ox\right)$ axis. This tilt modifies the value of all the coefficients $A_{n,m;n',m'}$. In particular, contribution from the Bragg spots which are not located on the diffraction row can be neglected and $A_{n,m;n',m'}\approx0$ if $m\neq0$  and/or $m'\neq0$. The two beam case is obtained after a second tilt of the sample around the $\left(Oy\right)$ axis. This tilt changes again the value of the coefficients $A_{n,m;n',m'}$ which become small except if $m=0$, $m'=0$, and small $n$, $n'$. In practice, this facilitates the numerical calculation of the essential Bloch wave coefficients. Eqs.~(\ref{eq:seven}) and (\ref{eq:eight}) show that $\left(\sigma_{2}-\sigma_{1}\right)$ is proportional to $\left(\mu_{+}-\mu_{-}\right)$. To express $\left(\sigma_{2}+\sigma_{1}\right)$ in a form which can further be used to derive the EMCD spin and orbital sum rules, we use the two additional approximations $\left(\mu_{+}+\mu_{-}\right)\approx\frac{2}{3}\left(\mu_{+}+\mu_{0}+\mu_{-}\right)$ and $\mu_{0}\approx\frac{1}{3}\left(\mu_{+}+\mu_{0}+\mu_{-}\right)$. Thanks to these approximations, $\left(\sigma_{2}+\sigma_{1}\right)$ becomes proportional to $\left(\mu_{+}+\mu_{0}+\mu_{-}\right)$.
The spin and orbital sum rules for an EMCD experiment can then be derived, using the sum rules which have been obtained by B. T. Thole et al. and P. Carra et al. to analyze XMCD spectra \cite{thole92,carra93}. The new EMCD sum rules can be written as
\begin{equation}
\frac{\int_{L_{3}}\left(\sigma_{2}-\sigma_{1}\right)dE-2\int_{L_{2}}\left(\sigma_{2}-\sigma_{1}\right)dE}{\int_{L_{3}+L_{2}}\left(\sigma_{2}+\sigma_{1}\right)dE}=K\left(\frac{2}{3}\frac{\left\langle S_{z}\right\rangle}{N_{h}}+\frac{7}{3}\frac{\left\langle T_{z}\right\rangle}{N_{h}}\right)\label{eq:nine}
\end{equation}
and
\begin{equation}
\frac{\int_{L_{3}+L_{2}}\left(\sigma_{2}-\sigma_{1}\right)dE}{\int_{L_{3}+L_{2}}
\left(\sigma_{2}+\sigma_{1}\right)dE}=K\frac{1}
{2}\frac{\left\langle L_{z}\right\rangle}{N_{h}}\label{eq:ten}
\end{equation}
where $\left\langle S_{z}\right\rangle/N_{h}$, $\left\langle L_{z}\right\rangle/N_{h}$ and $\left\langle T_{z}\right\rangle/N_{h}$ are respectively the ground state expectation values of spin momentum, orbital momentum, and magnetic dipole operators per hole in the $d$ bands. The coefficient $K$ contains all the information related to the dynamical effects. It can be expressed as
$$
K=3\sum_{\left(n,m\right)}\sum_{\left(n',m'\right)\neq\left(n,m\right)}\frac
{\textrm{Im}\left(A^{pos1}_{n,m;n',m'}\right)\left[\left(\delta-n\right)\left(\epsilon-
m'\right)-\left(\delta-n'\right)\left(\epsilon-m\right)\right]}{\left[\left
(\delta-n\right)^{2}+\left(\epsilon-m\right)^{2}+\frac{q^{2}_{z}}{g^{2}}\right]
\left[\left(\delta-n'\right)^{2}+\left(\epsilon-m'\right)^{2}+\frac{q^{2}_{z}}
{g^{2}}\right]}/
$$
$$
\{\sum_{\left(n,m\right)}\frac{A^{pos1}_{n,m;n,m}\left[\left(\delta-n\right)^{2}
+\left(\epsilon-m\right)^{2}+2\frac{q^{2}_{z}}{g^{2}}\right]}{\left[\left
(\delta-n\right)^{2}+\left(\epsilon-m\right)^{2}+\frac{q^{2}_{z}}{g^{2}}\right]
^2}
$$
\begin{equation}
+2\sum_{\left(n,m\right)}\sum_{\left(n',m'\right)\neq\left(n,m\right)}\frac
{\textrm{Re}\left(A^{pos1}_{n,m;n',m'}\right)\left[\left(\delta-n\right)\left(\delta-
n'\right)+\left(\epsilon-m\right)\left(\epsilon-m'\right)+2\frac{q^{2}_{z}}{g^
{2}}\right]}{\left[\left(\delta-n\right)^{2}+\left(\epsilon-m\right)^{2}+\frac
{q^{2}_{z}}{g^{2}}\right]\left[\left(\delta-n'\right)^{2}+\left(\epsilon-
m'\right)^{2}+\frac{q^{2}_{z}}{g^{2}}\right]}\}\label{eq:eleven}.
\end{equation}
$K$ can be calculated for a very well defined geometry. It will depend on the excitation error of the incident beam, the specimen thickness, the detector position and aperture size. Moreover, in the experiment one never can achieve a perfectly parallel beam. Convergence and partial coherence of the electron source make the precise calculation of $K$ untenable for the time being. Still, Eqs.~(\ref{eq:nine}) and (\ref{eq:ten}) can be used to obtain
\begin{equation}
\frac{\int_{L_{3}}\left(\sigma_{2}-\sigma_{1}\right)dE-2\int_{L_{2}}\left(\sigma_{2}-\sigma_{1}\right)dE}{\int_{L_{3}+L_{2}}
\left(\sigma_{2}-\sigma_{1}\right)dE}=\frac{4\left\langle S_{z}\right\rangle+14\left\langle T_{z}\right\rangle}{3
\left\langle L_{z}\right\rangle}\label{eq:twelve}
\end{equation}
free from any dynamical coefficient, sample orientation and thickness. Eqs.~(\ref{eq:nine}), (\ref{eq:ten}) and (\ref{eq:eleven}) apply to a single absorbing atom of the sample. The extension of the foregoing derivation from a four-fold symmetric diffraction pattern to the general case is straightforward.

\begin{figure}[h]
\begin{center}
\includegraphics[scale=1]{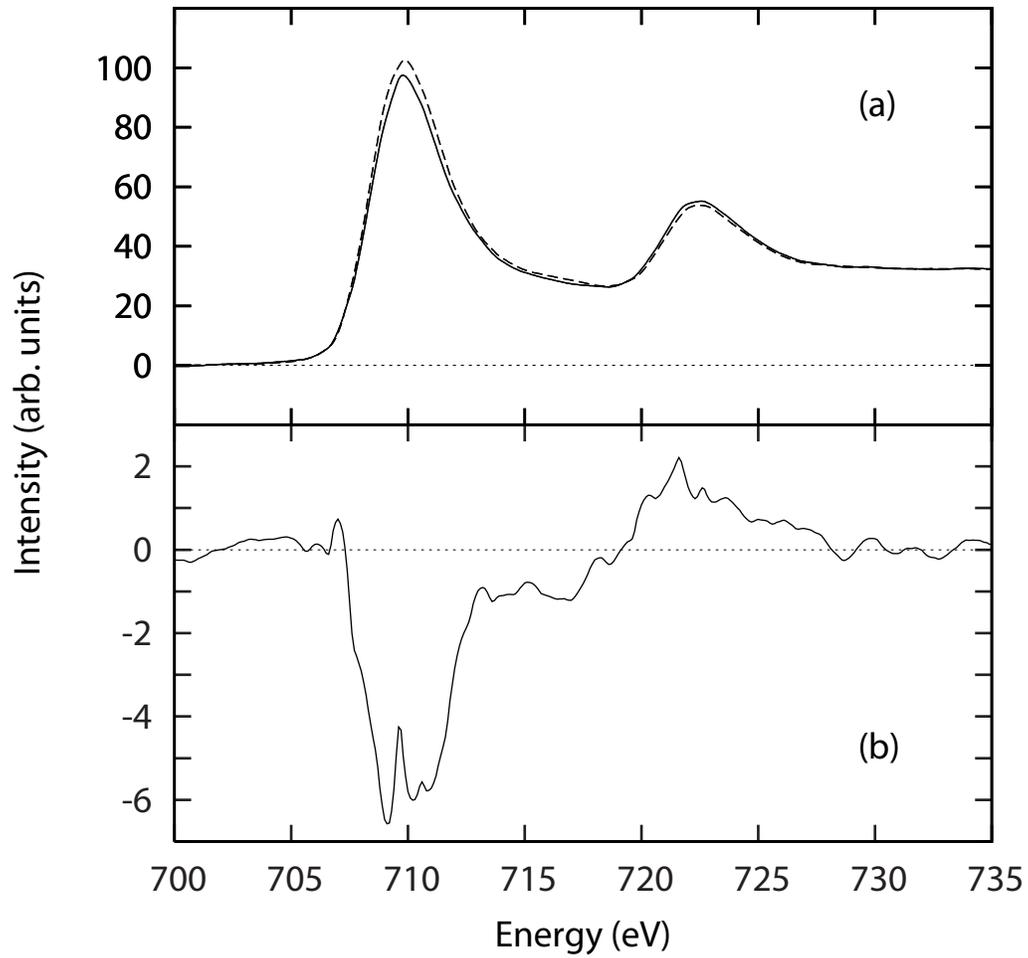}
\caption{\label{fig:spectra}(a): EELS signal measured at the two symmetric positions pos1 and pos2 in the diffraction pattern of an iron sample oriented in the (110) two beam configuration; (b): corresponding dichroic signal}
\end{center}
\end{figure}

We now briefly describe the experimental applicability of the EMCD sum rules. Experiments were performed using the SACTEM Toulouse, a TECNAI F20 (FEI) equipped with a spherical aberration corrector (CEOS), an Imaging Filter (Gatan Tridiem) and a 2k*2k Camera (Gatan). An iron sample was used as a test sample. By combining the techniques of tripod polishing and ion milling, we prepared a large flat area which was electron transparent. The magnetisation of the iron film is saturated in the $\left(Oz\right)$ direction by the field of the objective lens pole piece. The sample was oriented in (110) two beam configuration and the electron diffraction pattern was recorded using the ESI technique performed with a 1 eV slit in an energy range of [645 eV, 745 eV] for a total of 30 min exposure time \cite{warot07}. The diffraction pattern is taken using the LACDIF configuration \cite{midgley99,morniroli07,warot07} with a 7.8 mrad convergence angle which strongly increases the EMCD intensity and the signal/noise ratio. Post process corrections of isochromaticity and drift detected on the ESI data cube were applied using a home made software written in the scripting language of Digital Micrograph (Gatan). Finally, EELS spectra are extracted using the ESI data cube, for the two positions $\textbf{OS}=\frac{g}{2}\textbf{e}_{x}\pm\frac{g}{2}\textbf{e}_{y}$ located on the Thales circle which passes by the transmitted beam and the Bragg spot. Two circular apertures of semi-angle $\alpha=4.2$ mrad were used in the numerical integration, and the recorded spectra are shown in Fig.~\ref{fig:spectra}a.

The difference between the two spectra gives the dichroic signal which is represented in Fig.~\ref{fig:spectra}b. Our spectra have not been processed for removal of the background due to the 2$p$-state to continuum states transitions, because the aim of this paper is to demonstrate the feasability of the method. A quantitative analysis of spin/orbital moments would necessitate a better signal/noise ratio as well as more involved data treatment. We have applied Eq.~(\ref{eq:twelve}) to our experimental results, integrating the EMCD spectrum in the energy windows [705 eV, 715 eV] for the $L_{3}$ edge and [719 eV, 729 eV] for the $L_{2}$ edge. Neglecting the contribution of the magnetic dipole operator, this measurement has given $\left\langle L_{z}\right\rangle/\left\langle S_{z}\right\rangle=0.18\pm0.05$. This result is higher but with the same order of magnitude than the values 0.124 \cite{stearns86}, 0.088 \cite{bonnenberg86}, 0.133 \cite{carra93} and 0.086 \cite{chen95} which have been obtained from neutron scattering data, gyromagnetic ratio or XMCD spectra.

This comparison shows that EMCD is now on the way to giving quantitative magnetic information. Experiments do nevertheless deserve improvements, optimizing the angular and energy windows for integration in order to increase the still poor signal/noise ratio. Small background matching problems can also occur between the $L_{3}$ and $L_{2}$ edges. This can be seen near 715 eV where the dichroic signal does not perfectly vanish. These background problems are due to the fact that the non dichroic part of the signal is not perfectly the same at the two symmetric detector positions in the two beam case. In this case, Eq.~(\ref{eq:eight}) does not exactly describe the experimental configuration. This problem may be minimized by working at a higher voltage, in order to decrease the curvature of the Ewald sphere, or by looking for more symmetric experimental conditions for which Eq.~(\ref{eq:eight}) holds perfectly.

We have derived a set of sum rules for EMCD spectra which can be used to obtain orbital and spin moments of magnetic samples. Also dynamical diffraction effects of the electron beam in the specimen influence the dichroic spectra in a complicated way, the $\left\langle L_{z}\right\rangle/\left\langle S_{z}\right\rangle$ ratio can be extracted straightforwardly when the scattering conditions are properly chosen. The main advantage of using EELS instead of X-ray absorption for this quantitative analysis comes from the subnanometer probe size which can be reached in a TEM. This opens exciting perspectives for the local magnetic analysis of nanomaterials and nanodevices like magnetic tunnel junctions for spintronics applications or magnetic nanoparticles with enhanced anisotropy and magnetisation.

Acknowledments: The authors are grateful for support from the IP3 project of the 6th Framework Programme of the European Commission: ESTEEM (Enabling Science and Technology for European Electron Microscopy - Contract nr. 0260019), the French-Austria collaborative CNRS program (PHC-Amadeus). PS acknowledges support of the European commission, contract nr. 508971 (CHIRALTEM) and the CNRS Poste-Rouge program for funding his stay at CEMES-Toulouse.
%\end{acknowledgments}

%\bibliography{CALMELS}

\end{document}